\documentclass[a4paper,11pt]{amsart}
\usepackage{amsmath}
\usepackage{amstext,amsfonts,amsbsy,eucal,amssymb}

\usepackage{arrow}

\numberwithin{equation}{section}

\newtheorem{theorem}{Theorem}[section]
\newtheorem{lemma}[theorem]{Lemma}
\newtheorem{proposition}[theorem]{Proposition}

\newtheorem{conjecture}[theorem]{Conjecture}

\theoremstyle{definition}
\newtheorem{definition}[theorem]{Definition}
\newtheorem{example}[theorem]{Example}
\newtheorem{remark}[theorem]{Remark}

\newcommand{\e}{\operatorname{\mathrm{e}}}
\newcommand{\R}{\operatorname{\mathbb{R}}}
\newcommand{\Z}{\operatorname{\mathbb{Z}}}
\newcommand{\C}{\operatorname{\mathbb{C}}}
\newcommand{\I}{\operatorname{\mathbb{I}}}
\newcommand{\Div}{\operatorname{\mathrm{Div}}}
\newcommand{\Log}{\operatorname{\mathrm{Log}}}

\def\ba{\begin{array}}
\def\ea{\end{array}}
\def\pic{{\rm Pic}}

\def\noi{\noindent}
\def\nn{\nonumber}

\def\ve{\varepsilon}

\def\mr{{\mathbb R}}

\def\disp{\displaystyle}

\usepackage[dvipdfm]{graphicx,color}

\definecolor{spot}{cmyk}{1,0,0,0}

\begin{document}

\title{\large 
Tropical spectral curves and integrable cellular automata
}

\author{Rei Inoue}
\address{Department of Physics, Graduate School of Science,
The University of Tokyo,
7-3-1 Hongo, Bunkyo-ku, Tokyo 113-0033, Japan
\newline
CREST, JST, 4-1-8 Honcho Kawaguchi, Saitama 332-0012, Japan}
\email{reiiy@spin.phys.s.u-tokyo.ac.jp}

\author{Tomoyuki Takenawa} 
\address{Faculty of Marine Technology, 
Tokyo University of Marine Science and Technology,  
2-1-6 Etchu-jima, Koto-Ku, Tokyo, 135-8533, Japan}
\email{takenawa@kaiyodai.ac.jp }

\begin{abstract}
We propose a method to study 
the integrable cellular automata with periodic boundary conditions,
via the tropical spectral curve and its Jacobian.
We introduce the tropical version of eigenvector map
from the isolevel set to a divisor class on the tropical hyperelliptic curve.
We also provide some conjectures 
related to the divisor class and the Jacobian.
Finally, we apply our method to the periodic box and ball system and
clarify the algebro-geometrical meaning of the real torus introduced
for its initial value problem.
\end{abstract}


%\renewcommand{\subjclassname}{%
%  \textup{2000} Mathematics Subject Classification}
%\subjclass{Primary: 37J35. Secondary: 14H70, 14H40.}

\maketitle

%%%%%%%%%%%%%%%%%%%%%%
\section{Introduction}
%%%%%%%%%%%%%%%%%%%%%%%%%%%%%%%%%%%%
\subsection{Background and overview}
%%%%%%%%%%%%%%%%%%%%%%%%%%%%%%%%%%%%

The box and ball system (BBS) \cite{TakahashiSastuma90}
and the ultra-discrete Toda lattice \cite{TakaMatsu95}
are typical examples of integrable cellular automata on 
one-dimensional lattice.
The key to construct these systems from known soliton equations 
is a limiting procedure called {\it ultra-discretization} 
\cite{TTMS96}.
These automata are also well-defined on a periodic lattice,
and those are what we study in this paper.

In \cite{KimijimaTokihiro02}, Kimijima and Tokihiro
attempted to solve the initial value problem of 
the ultra-discrete periodic Toda lattice (UD-pToda). 
Their method consists of three steps: 
(1) send initial data of the UD-pToda to the discrete Toda lattice
 via inverse ultra-discretization, 
(2) solve the initial value problem for the discrete Toda
lattice and (3) take the ultra-discrete limit. 
However, due to technical difficulties, 
this method has been completed only in the case of genus $1$.
Thereafter the initial value problem of the pBBS is solved by a
combinatoric way \cite{MIT06} and by Bethe ansatz 
using Kerov-Kirillov-Reshetikhin bijection \cite{KTT06}.

In this paper we propose a method to study the isolevel set of 
the UD-pToda and the pBBS
via the tropical spectral curve and its Jacobian \cite{MikhaZhar06},
intending to solve the initial value problem. 
We introduce the tropical version of eigenvector map
from the isolevel set to a divisor class on the tropical hyperelliptic curve
(Propositions \ref{eigenvm-g=1}, \ref{eigenvm-g=2} etc).
We provide some conjectures 
(Conjectures \ref{conj:D-J} and \ref{conj:Toda-iso})
related to the divisor class and the Jacobian, 
and also present concrete computation in the case of genus $g\leq 3$.
Finally, by \eqref{BTJ-diagram} 
we unveil the algebro-geometrical meaning of the real torus
introduced in \cite{KTT06},   
on which the time evolution of the pBBS is linearized.

Tropical geometry is being established recently by many authors
(see \cite{Mikhalkin03,SpeyerSturm04} and references therein
for basic literature). 
It is defined over tropical semifield
$\mathbb{T}=\R\cup \{\infty\}$ 
equipped with the min-plus operation: 
$``x+y"=\min \{x,y\},\ ``xy"=x+y.$
In \cite{MikhaZhar06}, the Jacobian of a tropical curve
has been introduced by means of the corresponding metric graph.
Our approach might be a nice application
of tropical geometry to integrable systems
and one may confirm properness of the definition in \cite{MikhaZhar06}. 

%%%%%%%%%%%%%%%%%%%%%%%%%%%%%%%%%%%%%%%%
\subsection{Tropical curve and UD-pToda}
%%%%%%%%%%%%%%%%%%%%%%%%%%%%%%%%%%%%%%%%

We review on how tropical geometry appears in
studying the UD-pToda lattice.
Fix $g \in \Z_{>0}$.
The $(g+1)$-periodic Toda lattice of discrete time $t \in \Z$ 
\cite{HirotaTsujiImai} is given by the difference equations
on the phase space 
$\mathcal{U} = \{u^t = (I_1^t,\cdots,I_{g+1}^t, V_1^t,\cdots, V_{g+1}^t) 
~|~ t \in \Z \} \simeq \C^{2(g+1)}$: 
\begin{align}
  \label{d-Toda}
  I_i^{t+1} = I_i^t+V_i^t - V_{i-1}^{t+1},
  \qquad
  V_i^{t+1} = \frac{I_{i+1}^t V_i^t}{I_i^{t+1}},
\end{align}
where we assume the periodicity $I_{i+g+1}^t = I_i^t$ and 
$V_{i+g+1}^t = V_i^t$.
For each $u^t \in \mathcal{U}$, the Lax matrix is written as 
\begin{align}
  \label{d-Toda-Lax}
  L^t(y) = 
    \begin{pmatrix} 
      a_1^t & 1 & & & (-1)^g \frac{b_1^t}{y} \\
      b_2^t & a_2^t & 1 & & \\
      & \ddots & \ddots & \ddots & \\
      & & b_{g}^t & a_{g}^t & 1 \\
      (-1)^g y & & & b_{g+1}^t & a_{g+1}^t \\
    \end{pmatrix},
\end{align}
where $a_i^t = I_{i+1}^t+V_i^t$, $b_i^t = I_i^t V_i^t$
and $y \in \C$ is a spectral parameter.
The evolution \eqref{d-Toda} preserves $\det (x \I+L^t(y))$.
When we fix a polynomial $f(x,y) \in \mathbb{C}[x,y]$ as
\begin{align}
  \label{complex-curve}
  f(x,y) = y^2+y (x^{g+1}+c_{g} x^{g}+\cdots+c_1 x+c_0)+c_{-1},
\end{align}
the isolevel set $\mathcal{U}_c$ for \eqref{d-Toda} is 
$$
  \mathcal{U}_c = \{ u^t \in \mathcal{U} ~|~
  y \det (x \I+L^t(y)) = f(x,y) \}. 
$$
Let $\gamma_c$ be the algebraic curve given by $f(x,y) = 0$.
For generic $c_i$, $\gamma_c$ is the hyperelliptic curve of genus $g$.
Since the Lax matrix \eqref{d-Toda-Lax} is same as 
that for the original periodic Toda lattice (of continuous time) 
\cite{AdlerMoer80},  
$\mathcal{U}_c$ is isomorphic to the affine part of 
the Jacobi variety $\mathrm{Jac}(\gamma_c)$ of $\gamma_c$,
and the time evolution \eqref{d-Toda} is linearized on 
$\mathrm{Jac}(\gamma_c)$ \cite{KimijimaTokihiro02}. 

\begin{figure}
\begin{center}

\unitlength=1.2mm
\begin{picture}(100,80)(0,0)

\put(0,10){\line(1,0){20}}
\put(20,10){\vector(1,0){80}}
\put(20,0){\line(0,1){10}}
\put(20,10){\line(0,1){65}}
\put(20,75){\vector(0,1){5}}

\put(14,74){$C_{-1}$}
\put(16,6){$0$}
\put(28,10){\line(0,-1){1}}
\put(27,6){$\lambda_1$}
\put(36,10){\line(0,-1){1}}
\put(35,6){$\lambda_2$}
\put(45,6){$\cdots$}
\put(56,10){\line(0,-1){1}}
\put(55,6){$\lambda_{g-1}$}
\put(68,10){\line(0,-1){1}}
\put(67,6){$\lambda_{g}$}

\put(20,10){\line(-1,-2){5}}
\put(20,10){\line(4,5){8}}
\put(28,20){\line(1,1){8}}
\put(36,28){\line(2,1){10}}
\put(53,35){\line(5,1){15}}
\put(68,38){\line(1,0){10}}

\put(20,75){\line(-1,2){2}}
\put(20,75){\line(4,-5){8}}
\put(28,65){\line(1,-1){8}}
\put(36,57){\line(2,-1){10}}
\put(53,50){\line(5,-1){15}}
\put(68,47){\line(1,0){10}}

\put(28,20){\line(0,1){45}}
\put(36,28){\line(0,1){29}}
\put(56,35.5){\line(0,1){14}}
\put(68,38){\line(0,1){9}}

\end{picture}
\caption{Tropical hyperelliptic curve}\label{TropHyper}
\end{center}
\end{figure}
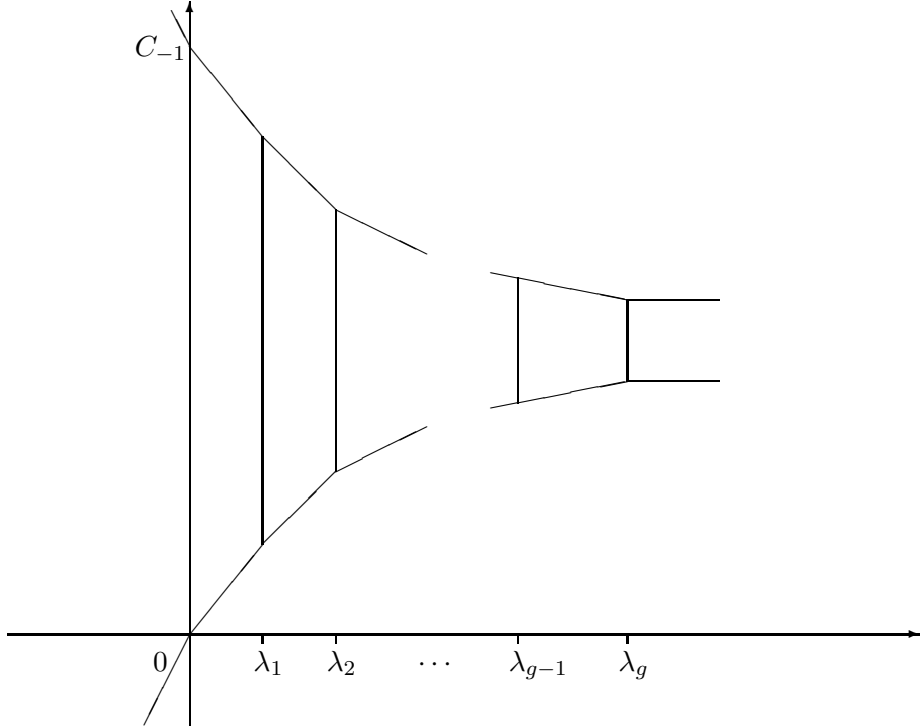

The {\it ultra-discrete limit} of \eqref{d-Toda}
is the UD-pToda 
\cite{NagaiTokihiroSatsuma98} given by the piecewise-linear map
$$T:\R^{2(g+1)}\to \R^{2(g+1)};\  
(Q_i^t,W_j^t) \mapsto (Q_i^{t+1},W_j^{t+1})$$
($t \in \Z$ and $i,j\in \{1,2,\dots,g+1\}$), where
\begin{align}
  \label{UD-pToda}
  \begin{split}
  &Q_i^{t+1} = \min[W_i^t, Q_i^t-X_i^t],
  \qquad 
  W_i^{t+1} = Q_{i+1}^t+W_i^t - Q_i^{t+1},
  \end{split}
\end{align}
with 
$X_i^t = \min_{k=0,\cdots,g}\bigl[\sum_{l=1}^k (W_{i-l}^t - Q_{i-l}^t)\bigr]$.
On the other hand, in this limit $\gamma_c$ is reduced 
to the tropical curve 
$\tilde{\Gamma}_C \subset \mathbb{R}^2$ given by
the polygonal lines of the convex in $\R^3$:
\begin{align}
  \label{trop-var}
  \{(X,Y,\min[2Y, (g+1)X+Y, gX+Y+C_g, \cdots, X+Y+C_1,Y+C_0,C_{-1}])\} 
\end{align}
For generic $C_i$ (see \eqref{CD-condition}), 
$\tilde{\Gamma}_C$ is {\it smooth} and depicted as Fig.~\ref{TropHyper}
where we fix $C_g = 0$ and 
set $\lambda_i = C_{g-i}-C_{g-i+1}$ for $i=1,\cdots,g$.
Note that all edges of $\tilde{\Gamma}$ have fractional slopes.

We explicitly construct a tropical version of the eigenvector map
from the isolevel set of the UD-pToda to the divisor class on
$\tilde{\Gamma}_C$, 
and show that the isolevel set
is isomorphic to the tropical Jacobi variety of $\tilde{\Gamma}_C$.

%%%%%%%%%%%%%%%%%%%%%%%%%%%%%%%%%%%%%%%%%%%%%%%%%%%%%%
\subsection{Ultra-discrete limit and min-plus algebra}
%%%%%%%%%%%%%%%%%%%%%%%%%%%%%%%%%%%%%%%%%%%%%%%%%%%%%%

We briefly introduce the notion of the ultra-discrete limit (UD-limit) 
and relate it to
the min-plus algebra 
on the tropical semifield $\mathbb{T} = \R \cup \{\infty\}$.

We define a map $\Log_\ve : \R_{>0} \to \R$ with an infinitesimal parameter 
$\ve > 0$ by
\begin{align}
  \label{loge-map}
  \Log_\ve : x \mapsto - \ve \log x.
\end{align}
For $x > 0$, we define $X \in \mathbb{T}$ by $x = \e^{-\frac{X}{\ve}}$.
Then the limit $\ve \to 0$ of $\Log_\ve (x)$ converges to $X$.    
The procedure $\lim_{\ve \to 0} \Log_\ve$ with the scale transformation as 
$x = \e^{-\frac{X}{\ve}}$
is called the ultra-discrete limit. 

We summarize this procedure in more general setting:
\begin{proposition}
For $A,B,C\in \mr$ and $k_a,k_b,k_c>0$, 
set 
$$a=k_a e^{-\frac{A}{\ve}},\ b=k_b e^{-\frac{B}{\ve}}, 
\ c=k_c e^{-\frac{C}{\ve}}$$
and take the limit $\ve \to 0$ of the image $\Log_\ve$ of 
%$\disp -\lim_{\ve\to 0}\ve \log(.)$ of 
the equations
$$({\rm i})\ a+b=c,\ ({\rm ii})\ ab=c,\ ({\rm iii})\ a-b=c.$$
Then 
$$({\rm i})\ \min [A,B]=C,\ ({\rm ii})\ A+B=C$$
and
$$({\rm iii}) 
\left\{
\ba{ll} A=C& \ \ ({\rm if} \ \ A<B, ~{\rm or }~ A=B~ {\rm and }~ k_a > k_b)\\
{\rm contradiction}& \ \ ({\rm otherwise})
\ea\right.$$
hold.
\end{proposition}

%%%%%%%%%%%%%%%%%%%%
\subsection{Content}
%%%%%%%%%%%%%%%%%%%%

In \S 2,
we define the metric graph $\Gamma_C$ 
for the tropical hyperelliptic curve 
$\tilde{\Gamma}_C$ and define its Jacobian $J(\Gamma_C)$.
By using a tropical version of the Abel-Jacobi map,
we propose a divisor class which isomorphic to $J(\Gamma_C)$
at Conjecture \ref{conj:D-J}. 
This is justified for $g \leq 3$. 
In \S 3, we study the isolevel set of the UD-pToda.
We construct the eigenvector map
from the isolevel set to the divisor class on the tropical curve.
It is shown that the general level set is isomorphic to $J(\Gamma_C)$.
In \S 4, we clarify the correspondence of the UD-pToda with the pBBS
by refining that in \cite{KimijimaTokihiro02}.
In conclusion we interpret the isolevel set of the pBBS 
introduced in \cite{KTT06} in terms of tropical geometry.

%%%%%%%%%%%%%%%%%%%%%%%%%%%%%
\subsection*{Acknowledgement}
%%%%%%%%%%%%%%%%%%%%%%%%%%%%%

R.~I. thanks Atsuo Kuniba for discussion.
She also thanks Keigo Hijii for quick help in using LaTeX.
T.~T. appreciates the assistance from the 
Japan Society for the Promotion of Science.
R.~I is supported by Grant-in-Aid for Young Scientists (B) (19740231).

%%%%%%%%%%%%%%%%%%%%%%%%%%%%%%%%%%%%%%%%%%%%%%%%%%%
\section{Tropical hyperelliptic curve and Jacobian}
%%%%%%%%%%%%%%%%%%%%%%%%%%%%%%%%%%%%%%%%%%%%%%%%%%%

\subsection{Tropical hyperelliptic curve}

Fix $g \in \Z_{>0}$ and $C=(C_{-1},C_0,\cdots,C_g) \in \R^{g+2}$.
Let $\tilde{\Gamma}_C \subset \mathbb{R}^2$ 
be the affine tropical curve given by the polygonal lines of the convex 
in $\R^3$ \eqref{trop-var}.
We assume a generic condition for $C$: 
\begin{align}
  \label{CD-condition}  
  C_{-1} > 2 C_0, ~  
  C_i+C_{i+2} > 2 C_{i+1} ~ (i=0,\cdots,g-2),~
  C_{g-1} > 2 C_g.
\end{align}
For simplicity, we fix $C_g = 0$ in the following.
Define $\lambda =(\lambda_1,\cdots,\lambda_g)$ and 
$p_1,\cdots,p_g$ by 
\begin{align}
  \label{partition}
  \lambda_i = C_{g-i}-C_{g-i+1}, \qquad
  p_i = C_{-1} - 2 \sum_{j=1}^g \min[\lambda_i,\lambda_j]. 
\end{align}
Under the condition \eqref{CD-condition} one sees 
$0 < \lambda_1 < \lambda_2 < \cdots < \lambda_g$ and 
$2 \sum_{i=1}^g \lambda_i < C_{-1}$. 

By referring \cite[Definition 2.18]{Mikhalkin03},
we introduce a notion of smoothness of tropical curves:
\begin{definition}
  The tropical curve $\Sigma \hookrightarrow \R^2$ is smooth 
  if the following conditions are satisfied:
  \\
  (a) all edges in $\Sigma$ have fractional slopes.
  \\
  (b) All vertex $v \in \Sigma$ is $3$-valent.
  \\
  (c) For each $3$-valent vertex $v$,  
  let $e_1,e_2,e_3$ be the oriented edges outgoing from $v$.  
  Then the primitive tangent vectors $\xi_k$ of $e_k$  
  satisfy $\sum_{k=1}^3 \xi_k = 0,$ and $|\xi_k \wedge \xi_j| = 1$
  for $k \neq j$, $k,j \in \{1,2,3\}$.
\end{definition}
We see that $\tilde{\Gamma}_C$ is smooth.
In particular, it is a tropical hyperelliptic curve whose 
genus is $\dim H_1(\tilde{\Gamma}_C,\Z) = g$
(see Fig.~\ref{TropHyper}).
We are to consider the maximal compact subset 
$\Gamma_C = \tilde{\Gamma}_C \setminus \{\text{infinite edges}\}$
of $\tilde{\Gamma}_C$.
For simplicity we write $\Gamma$ for $\Gamma_C$.

\subsection{Metric on $\Gamma$}

Following \cite[\S 3.3]{MikhaZhar06}, we equip $\Gamma$ with the structure of 
a metric graph. 
Let $\mathcal{E}(\Gamma)$ be the set of edges in $\Gamma$,
and define the weight $w : \mathcal{E}(\Gamma) \to \R_{\geq 0}$ by
$$
  w(e) = \frac{\parallel e \parallel}{\parallel \xi_e \parallel},
$$
where $\xi_e$ is the primitive tangent vector of $e \in \mathcal{E}(\Gamma)$, 
and $\parallel ~ \parallel$ denotes any norm in $\R^2$.
With this weight 
the tropical curve $\Gamma$ becomes a metric graph.

\begin{figure}
\begin{center}
\unitlength=1.2mm
\begin{picture}(100,70)(0,10)

\put(25,13){$\lambda_1$}
\put(33,22){$\lambda_2-\lambda_1$}
\put(58,33){$\lambda_g-\lambda_{g-1}$}

\put(25,69){$\lambda_1$}
\put(32,62){$\lambda_2-\lambda_1$}
\put(58,50){$\lambda_g-\lambda_{g-1}$}

\put(20,10){\line(4,5){8}}
\put(28,20){\line(1,1){8}}
\put(36,28){\line(2,1){10}}
\put(53,35){\line(5,1){15}}

\put(20,75){\line(4,-5){8}}
\put(28,65){\line(1,-1){8}}
\put(36,57){\line(2,-1){10}}
\put(53,50){\line(5,-1){15}}

\put(20,10){\line(0,1){65}}
\put(28,20){\line(0,1){45}}
\put(36,28){\line(0,1){29}}
\put(56,35.5){\line(0,1){14}}
\put(68,38){\line(0,1){9}}

\put(18.5,43){$C_{-1}$}
\put(27,43){$p_1$}
\put(35,43){$p_2$}
\put(43,43){$\cdots$}
\put(51,43){$p_{g-1}$}
\put(67,43){$p_{g}$}

\put(24,42){\oval(4,45)}
\put(26,30){\vector(0,1){3}}
\put(22.5,30){$\alpha_1$}
\put(32,42){\oval(4,27)}
\put(34,35){\vector(0,1){3}}
\put(30.5,35){$\alpha_2$}
\put(62,42){\oval(8,8)}
\put(66,41){\vector(0,1){3}}
\put(61.5,41){$\alpha_g$}

\end{picture}
\caption{$\Gamma_C$ as a metric graph}\label{Gamma-metric}
\end{center}
\end{figure}
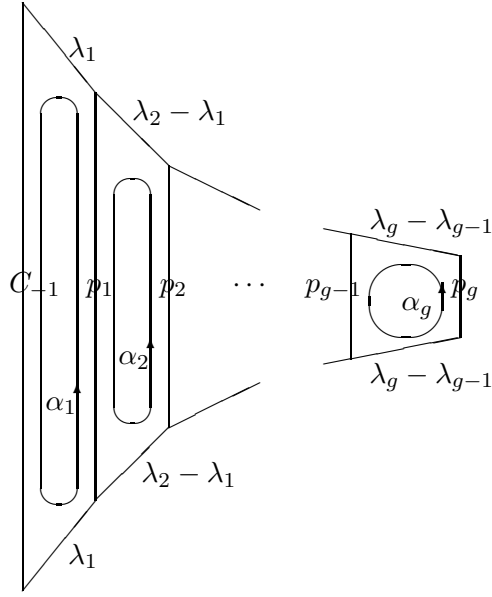

The metric on $\Gamma$ defines a symmetric bilinear form $Q$ 
on the space of paths in $\Gamma$ as follows:
for a non-self-intersecting path $\gamma$,
set $Q(\gamma, \gamma) := \mathrm{length}_w(\gamma)$,
and extending it to any pairs of paths bilinearly.
In Fig.~\ref{Gamma-metric}
we show the weight for each edge in $\Gamma$ and 
the basis $\alpha_i ~(i=1,\cdots,g)$ of $\pi_1(\Gamma)$.
For example, we have $Q(\alpha_1,\alpha_1) = C_{-1}+p_1+2 \lambda_1$,
$Q(\alpha_1,\alpha_2) = -p_1$, and 
$Q(\alpha_1,\alpha_i) = 0$ for $i>2$.

\subsection{Tropical Jacobian}

Let $\Omega(\Gamma)$ be the space of global $1$-forms on $\Gamma$,
and $\Omega(\Gamma)^\ast$ be the dual space of $\Omega(\Gamma)$.
Then both $\Omega(\Gamma)$ and $\Omega(\Gamma)^\ast$ are $g$ dimensional
and $\Omega(\Gamma)^\ast$ is isomorphic to $H_1(\Gamma,\R)$.

\begin{definition}\cite[\S 6.1]{MikhaZhar06}
  The tropical Jacobian of $\Gamma$ is a $g$ dimensional real torus 
  defined as
  $$ 
    J(\Gamma) = \Omega(\Gamma)^\ast / H_1(\Gamma,\Z)
    \simeq \R^g / K \Z^g \simeq \R^g / \Lambda \Z^g.
  $$
  Here $K, \Lambda \in M_g(\R)$ are given by
  \begin{align*}
    K_{ij} =& Q(\alpha_i, \alpha_j), \\
    \Lambda_{ij} =& Q(\sum_{k=1}^i \alpha_k,\sum_{l=1}^j \alpha_l)
                 = C_{-1}+p_i \delta_{ij}+2 \min[\lambda_i,\lambda_j].
  \end{align*}
\end{definition}
Since $Q$ is nondegenerate,
$K$ and $\Lambda$ are symmetric and positive definite.
In particular, we say that $J(\Gamma)$ is principally polarized.

Let $\Div_{\mathrm{eff}}^g(\Gamma)$ be a set of 
effective divisors of degree $g$ on $\Gamma$.
We fix $P_0 \in \Gamma$ and define a map 
$\eta :~ \Div_{\mathrm{eff}}^g(\Gamma) \to J(\Gamma)$;
\begin{align}
  \label{UD-AJ}
  P_1+\cdots+ P_g \mapsto 
  \sum_{i=1}^g (Q(\gamma_i,\alpha_1),\cdots,Q(\gamma_i,\alpha_g)),
\end{align}
where $\gamma_i$ is the path from $P_0$ to $P_i$ on $\Gamma$.
Define 
$\alpha_{ij} = \alpha_i \cap \alpha_j \setminus 
\{\text{the end-points of $\alpha_i \cap \alpha_j$}\} \subset \Gamma$,
and $\mathcal{D}^g(\Gamma)$ to be a subset of $\Div_{\mathrm{eff}}^g(\Gamma)$: 
\begin{align*}
  \mathcal{D}^g(\Gamma) 
  = \Bigl\{P_1+\cdots+P_g ~\Big|~
       \begin{array}{l}
        \text{$P_i \in \alpha_i$ for all $i$, and} \\ 
        \text{there exists at most one point on $\alpha_{ij}$ for all 
        $i \neq j$}
        \end{array} 
        \Bigr\}
\end{align*}                         

\begin{conjecture}\label{conj:D-J}
  A reduced map $\eta |_{\mathcal{D}^g(\Gamma)}$ is bijective:
  $$ 
  \eta |_{\mathcal{D}^g(\Gamma)}: ~ 
    \mathcal{D}^g(\Gamma) \stackrel{\sim}{\to} J(\Gamma).
  $$
\end{conjecture}
In the case of $g=1$, this conjecture is obviously true 
since $\mathcal{D}^g(\Gamma) = \Gamma \simeq J(\Gamma)$ by definition. 
In the following we show that this conjecture is true for $g=2$ and $3$.
\begin{proof}
  We define a map 
  $\iota_S :~ \Gamma \to \R^g ; 
           ~ P \mapsto \iota_S(P) = (Q(\gamma,\alpha_i))_{1 \leq i \leq g}$
  where $S \in \Gamma$ and $\gamma$ is an appropriate path 
  from $S$ to $P$.
  For $P_1+\cdots+P_g \in \Div_{\mathrm{eff}}^g(\Gamma)$,
  we see $\eta(P_1+\cdots+P_g) \sim \sum_{i=1}^g \iota_{P_0}(P_i)$
  in $J(\Gamma)$.
  
  $g=2$ case:
We set $P_0 = (\lambda_1, 2 \lambda_1)$ which is 
the end-point of $\alpha_{12}$.
In the left figure of Fig.~\ref{Etag=2}
we illustrate the locus of $\iota_{P_0}(P_i)$
where $P_i$ starts from $P_0$ and moves along $\alpha_i$ for $i=1,2$ 
respectively.
We set $O = (0,0), A_1 = (C_{-1}+ p_1+2 \lambda_1, -p_1)$ 
and $A_2 = (-p_1, 2 p_1)$. 
The parallelogram $F$ of dash lines
is the fundamental domain of $J(\Gamma)$.
We calculate the image of the map 
$\mathcal{D}^2(\Gamma) \to \R^2$ given by  
$P_1+P_2 \mapsto \iota_{P_0}(P_1)+\iota_{P_0}(P_2) 
+ \overrightarrow{A_2 O}$,
and obtain the parallelohexagon $V$ composed of 
three non-overlapped parallelograms as shown in the right figure
of Fig.~\ref{Etag=2}. 
It is easy to see that $V$ is isomorphic to $F$ in $J(\Gamma)$.
\begin{figure}   
\begin{center}
\unitlength=1mm
\begin{picture}(140,50)(-5,0)

\put(0,30){\vector(1,0){55}}
\put(10,0){\vector(0,1){60}}

%\put(20,30){\line(0,1){1}}
%\put(45,30){\line(0,1){1}}
%\put(10,50){\line(1,0){1}}

\put(10,30){\line(1,0){1}}

\thicklines
\put(10,30){\line(1,-1){10}}
\put(20,20){\line(1,0){25}}
\put(10,30){\line(0,1){10}}
\put(10,40){\line(-1,1){10}}

\put(44.3,19){$\bullet$} 
\put(47,20){$A_1$}
\put(30,16){$\iota_{P_0}(P_1)$}
\put(-1,49){$\bullet$}
\put(-3,52){$A_2$}
\put(-9,39){$\iota_{P_0}(P_2)$}

\put(6,26){$O$}
%\put(12,39){$p_1$}
%\put(12,49){$2 p_1$}
%\put(19,33){$p_1$}
%\put(40,33){$C_{-1}+p_1+2\lambda_1$}

\multiput(10,30)(3.5,-1){11}{\line(1,0){0.8}}
\multiput(0,50)(3.5,-1){10}{\line(1,0){0.8}}
\multiput(0,50)(1,-2){11}{\line(0,1){0.5}}
\multiput(35,40)(1,-2){11}{\line(0,1){0.5}}

%%%%%%%

\thinlines

\put(70,30){\vector(1,0){55}}
\put(80,0){\vector(0,1){60}}

\put(90,30){\line(0,1){1}}
\put(115,30){\line(0,1){1}}
\put(80,50){\line(1,0){1}}
\put(80,20){\line(-1,0){1}}
\put(80,10){\line(-1,0){1}}

\put(80,30){\line(1,0){1}}

\thicklines
\put(80,30){\line(1,-1){10}}
\put(90,20){\line(1,0){25}}
\put(80,30){\line(0,1){10}}
\put(80,40){\line(-1,1){10}}

\put(100,0){\line(0,1){10}}
\put(100,10){\line(-1,1){10}}
\put(125,0){\line(0,1){10}}
\put(125,10){\line(-1,1){10}}
\put(90,10){\line(0,1){10}}
\put(100,0){\line(-1,1){10}}
\put(100,10){\line(1,0){25}}
\put(100,0){\line(1,0){25}}

\multiput(90,10)(3.5,-1){10}{\line(1,0){0.8}}
\multiput(80,30)(3.5,-1){10}{\line(1,0){0.8}}
\multiput(80,30)(1,-2){11}{\line(0,1){0.5}}
\multiput(115,20)(1,-2){11}{\line(0,1){0.5}}

\put(76,26){$O$}
\put(82,39){$p_1$}
\put(82,49){$2 p_1$}
\put(70,19){$-p_1$}
\put(70,9){$-2 p_1$}

\put(89,33){$p_1$}
\put(110,33){$C_{-1}+p_1+2\lambda_1$}

\end{picture}
\caption{Imege of $\eta$ in $g=2$}\label{Etag=2}
\end{center}
\end{figure}
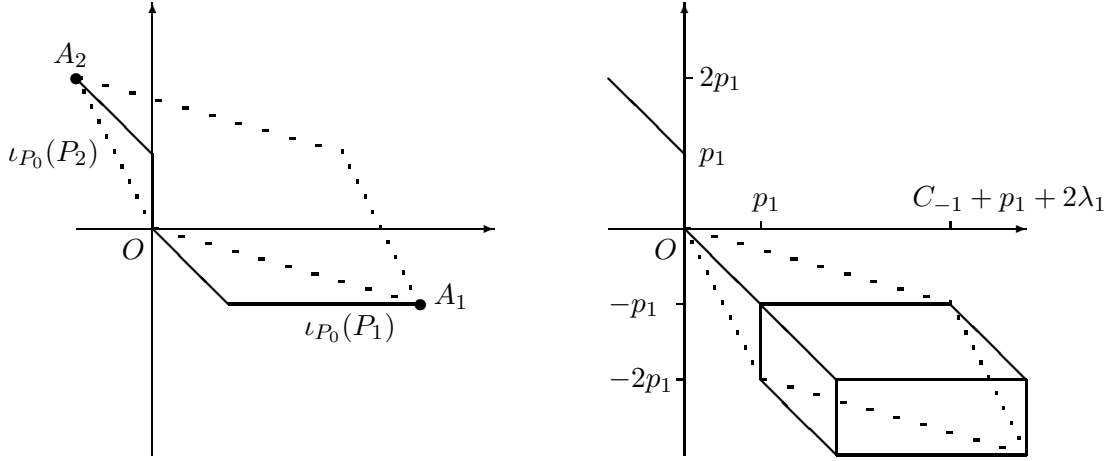

  $g=3$ case:
  We set $S_1 = (\lambda_1, 3 \lambda_1)$ and 
  $S_2 = (\lambda_2, \lambda_1+2 \lambda_2)$,
  the end-points of $\alpha_{12}$ and $\alpha_{23}$ respectively.  
  We calculate the image of the map
  $\mathcal{D}^3(\Gamma) \to \R^3$ given by 
  $P_1+P_2+P_3 \mapsto 
  \iota_{S_1}(P_1)+\iota_{S_1}(P_2)+ \iota_{S_2}(P_3)$,
  and obtain non-overlapped $12$ parallelopipeds.
  After shifting some parallelopipeds 
  along the lattice $K \Z^3$,
  we obtain the parallelo dodecahedron $V$ in Fig.~\ref{Etag=3}. 
We set $O = (0,0,0), ~A_1 = (C_{-1}+p_1+2\lambda_1,-p_1,0), ~
A_2 = (p_1,-p_1-p_2-2(\lambda_2-\lambda_2),p_2), ~A_3 = (0,-p_2,2 p_2)$,
$P = A_1+A_2 +A_3$
and $B_i = A_j+A_k$ for $\{i,j,k\} = \{1,2,3\}$.
The parallelopiped $F$ spanned by 
$\overrightarrow{O A_1}, \overrightarrow{O A_2}$ and 
$\overrightarrow{O A_3}$ 
is the fundamental domain of $J(\Gamma)$.
We draw $V$ in black, and $F$ in blue. 

\begin{figure}
\begin{center}
\unitlength=1mm
\begin{picture}(100,100)(-10,0)

\put(0,80){\vector(1,0){90}} \put(92,80){$z_1$}
\put(20,0){\line(0,1){29}}
\put(20,80){\vector(0,1){10}} \put(19,92){$z_2$}
\put(35,90){\vector(-3,-2){50}} \put(-20,55){$z_3$}

\thicklines
\put(20,80){\line(1,0){40}}
\put(60,80){\line(1,-1){15}}
\put(11,74){\line(1,0){40}}
\put(51,74){\line(1,-1){15}}
\put(2,59){\line(1,0){40}}
\put(42,59){\line(1,-1){15}}

\put(20,80){\line(-3,-2){9}}
\put(11,74){\line(-3,-5){9}}
\put(60,80){\line(-3,-2){9}}
\put(51,74){\line(-3,-5){9}}
\put(75,65){\line(-3,-2){9}}
\put(66,59){\line(-3,-5){9}}

\put(2,59){\line(0,-1){15}}
\put(2,44){\line(1,-1){15}}
\put(17,29){\line(1,0){40}}
\put(57,29){\line(0,1){15}}

\put(57,29){\line(3,2){9}}
\put(66,35){\line(3,5){9}}
\put(75,50){\line(0,1){15}}

\multiput(17,29)(3,2){4}{\line(0,1){0.5}}
\multiput(26,35)(1.5,2.5){7}{\line(0,1){0.5}}
\multiput(2,44)(3,2){4}{\line(0,1){0.5}}
\multiput(11,50)(1.5,2.5){7}{\line(0,1){0.5}}

\multiput(11,50)(2,-2){8}{\line(0,1){0.5}}
\multiput(26,35)(4,0){10}{\line(1,0){0.5}}
\multiput(20,80)(0,-2.5){6}{\line(0,1){0.5}}
\multiput(20,65)(2,-2){8}{\line(0,1){0.5}}
\multiput(35,50)(4,0){10}{\line(1,0){0.5}}

\put(19,79){$\bullet$}\put(16,82){$O$}
\put(74,64){$\bullet$}\put(76,66){$A_1$}
\put(1,58){$\bullet$}\put(-4,58){$A_3$}
\put(25,34){$\bullet$}\put(19,35){$A_2$}

\put(6,12){$\bullet$}\put(1,12){$B_1$}
\put(56,43){$\bullet$}\put(59,43){$B_2$}
\put(80,18){$\bullet$}\put(82,18){$B_3$}
\put(61,-3){$\bullet$}\put(64,-4){$P$}

\put(11,62){$\ast^\prime$}
\put(66,47){$\ast$}
\put(41,76){$\star$}
\put(47,31){$\star^\prime$}
\put(23,43){$\diamond$}
\put(41,64){$\diamond^\prime$}

%% fundamental domain %%%
\color{spot}
\multiput(20,80)(2.2,-0.6){26}{\line(0,1){1.5}}
\multiput(20,80)(0.3,-2.7){17}{\line(1,0){0.5}}
\multiput(20,80)(-1.5,-1.75){13}{\line(0,1){1.5}}

\multiput(2,58)(2.2,-0.6){26}{\line(0,1){1.5}}
\multiput(2,58)(0.3,-2.7){17}{\line(1,0){1.5}}
\multiput(57,42.6)(0.3,-2.7){17}{\line(1,0){1.5}}
\multiput(75,63.6)(0.3,-2.7){17}{\line(1,0){1.5}}
\multiput(75,63.6)(-1.5,-1.75){13}{\line(0,1){1.5}}
\multiput(80.4,18)(-1.5,-1.75){13}{\line(0,1){1.5}}
\multiput(7,12)(2.2,-0.6){26}{\line(0,1){1.5}}
\multiput(7,12)(1.5,1.75){13}{\line(0,1){0.5}}
\multiput(25,34)(2.2,-0.6){26}{\line(0,1){0.5}}

%%%%%%%%%%%%%%%

\end{picture}
\caption{Image of $\eta$ in $g=3$}\label{Etag=3}
\end{center}
\end{figure}
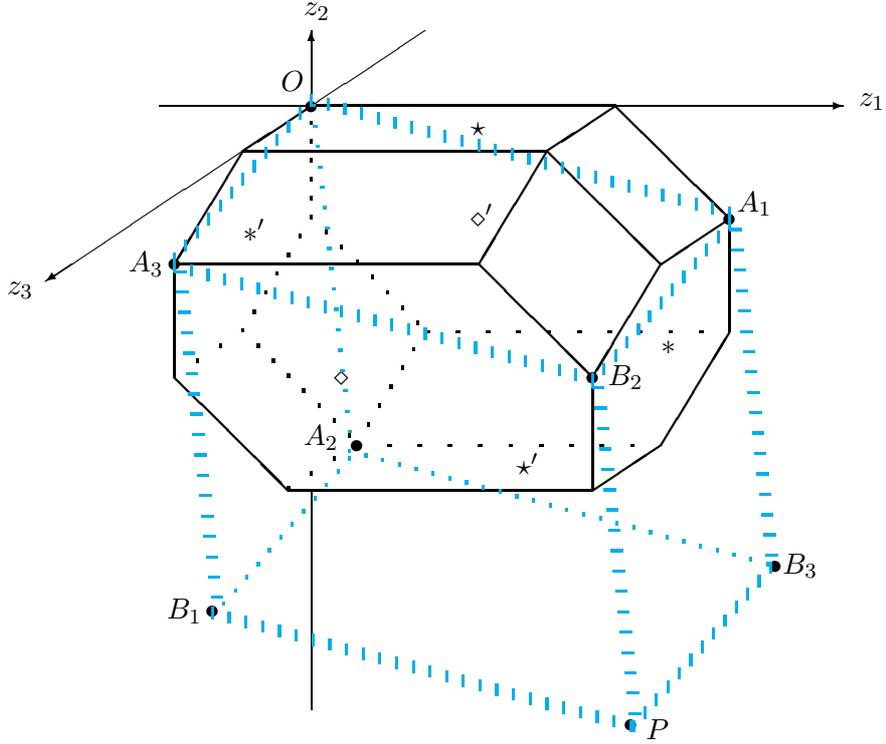

One sees that $V$ coincides with $F$ in $J(\Gamma)$ as follows:
The polygon $V \setminus F$ is composed of three parts
each of which contains the face $\ast^\prime$ in $z_2 z_3$-plane,
the face $\star$ in $z_1 z_3$-plane or 
the face $\diamond^\prime$ in $z_1 z_2$-plane.
We translate the part with the face $\ast^\prime$
(resp. $\star$, $\diamond^\prime$)  
by $\overrightarrow{O A_1}$ (resp. $\overrightarrow{O A_2}$,
$\overrightarrow{O A_3}$)
and attach it on the face $\ast$ (resp. $\star^\prime$, $\diamond$).    
\end{proof}

\begin{remark}
  After this paper was submitted, 
  we proved Conjecture \ref{conj:D-J} for general $g$ 
  in another way, by applying the notion of rational functions on $\Gamma$
  \cite{GathKerber06,MikhaZhar06}.
\end{remark}

For \S~4.2, we introduce another torus $J'(\Gamma)$: 
\begin{align}
  \label{Mat-T}
  J'(\Gamma) = \R^g / A \Z^g,
\end{align}
where $A_{ij} = \Lambda_{ij} - C_{-1}$. 

\begin{lemma}\label{lemma:Lambda-A}
  (i) $\det \Lambda = \det K = 
       (g+1) \det A = (g+1)  p_1 \cdots p_{g-1}C_{-1}$.
  \\
  (ii) Let $\nu_{\Gamma}$ be a shift operator, 
       $\nu_{\Gamma} :~ \R^g \to \R^g;
       ~ (z_i)_{i=1,\cdots,g} \mapsto (z_i+C_{-1})_{i=1,\cdots,g}$. 
%       Then $J(\Gamma) / (\nu_{\Gamma} \cdot \Z^g) = J'(\Gamma)$.
        Then $J'(\Gamma) \simeq 
              J(\Gamma) / \{P \sim \nu_{\Gamma}(P) ~|~ P \in J(\Gamma) \}$. 
\end{lemma}
The proof is elementary and left for readers.

%%%%%%%%%%%%%%%%%%%%%%%%%%%%%%%%%%%%%%%%%%%%%%%%%%%%%%
\section{Isolevel set of ultra-discrete periodic Toda}
%%%%%%%%%%%%%%%%%%%%%%%%%%%%%%%%%%%%%%%%%%%%%%%%%%%%%%

\subsection{Periodic Toda lattice}

We review the known results on the ($g+1$)-periodic Toda lattice 
\eqref{d-Toda}.
We define a matrix $M^t(y)$ besides the Lax matrix $L^t(y)$ \eqref{d-Toda-Lax}
on the phase space $\mathcal{U}$:
$$
  M^t(y) = 
    \begin{pmatrix} 
      I_2^t & 1 & & &  \\
       & I_3^t & 1 & & \\
      & & \ddots & \ddots & \\
      & &  & I_{g+1}^t & 1 \\
      y& & & & I_1^t \\
    \end{pmatrix}.
$$
\begin{proposition}\cite{HirotaTsujiImai}
  (i)
  The system \eqref{d-Toda} is equivalent to 
  the Lax form 
  $$ L^{t+1}(y)M^t(y)=M^t(y)L^t(y).$$
  (ii)
  The system \eqref{d-Toda} preserves the characteristic polynomial 
  of $L^t(y)$, $\det(x \I+L^t(y))$. 
\end{proposition}
\begin{proof}
(i)
Set $R^t(y)$ as
\begin{align*}
  R^t(y) = 
    \begin{pmatrix} 
      1 &  &  & (-1)^{g}\frac{V_1^t}{y} \\
      V_2^t & 1   & & \\
      & \ddots & \ddots  & \\
      & & V_{g+1}^t & 1 \\
    \end{pmatrix}.
\end{align*}
The system (\ref{d-Toda}) is equivalent to 
$R^{t+1}(y)M^{t+1}(y)=M^t(y)R^t(y)$. 
By the fact $L^t(y)=R^t(y)M^t(y)$, we have 
$$L^{t+1}(y)M^t(y)=R^{t+1}(y)M^{t+1}(y)M^t(y)
=M^t(y)R^t(y)M^t(y)=L^t(y)M^t(y).$$ 
(ii)
From the Lax form  we obtain 
$\det(x \I+L^{t+1}(y))=\det(x \I+M^t(y)L^t(y)(M^t(y))^{-1})
 =\det(x \I+L^{t}(y)).$
\end{proof}

We define the (complex) spectral curve $\gamma_c$ given by 
\begin{align}\label{spectralcurve} 
  \begin{split}
    f(x,y) 
    =& y\det(\I x+L^t(y)) \\
    =& y^2+y (x^{g+1}+c_{g} x^{g}+\cdots +c_0) +c_{-1} = 0.
  \end{split}
\end{align}
Concretely, $c_i$ is given by 
(for simplicity, we write $I_i^t=I_i,V_i^t=V_i$ and so on)
\begin{align}
  \begin{split}
  &c_{g} = \sum_{1\leq i \leq g+1} I_i+\sum_{1\leq i \leq g+1} V_i,
    \\ \label{C_g-1}
  &c_{g-1} = \sum_{1\leq i<j \leq g+1} (I_i I_j)
                  +\sum_{1\leq i<j \leq g+1} (V_i V_j)
                  +\sum_{1\leq i,j\leq g+1, j \neq i,i-1} (I_i V_j),
  \\ 
  & \vdots
  \\
  &c_0 = \prod_{i=1}^{g+1} I_i+ \prod_{i=1}^{g+1} V_i,
  \\
  &c_{-1}= \prod_{i=1}^{g+1} I_i V_i.
  \end{split} 
\end{align}
For generic $c_i$, $\gamma_c$ is a hyperelliptic curve.
Since \eqref{d-Toda} is invariant under 
$(I_i,V_i)_{1 \leq i \leq g+1} \mapsto 
(I_i c_g, V_i c_g)_{1 \leq i \leq g+1}$,
we can set $c_g = 1$ without loss of generality.

\begin{proposition}\cite{KimijimaTokihiro02}
  Under the condition
  $\prod_{k=1}^{g+1}V_k^t \neq \prod_{k=1}^{g+1}I_k^t$,
  the system \eqref{d-Toda} is equivalent to the system: 
  \begin{align}\label{d-toda2}
  \ba{rcl}
  I_i^{t+1}&=&\disp V_i^t+I_i^t
  \frac{1-\prod_{k=1}^{g+1}\frac{V_k^t}{I_k^t}}{
  1+\sum_{j=1}^g\prod_{k=1}^{j}\frac{V_{i-k}^t}{I_{i-k}^t}},
  \\
  V_i^{t+1}&=&\disp \frac{I_{i+1}^tV_i^t}{I_i^{t+1}}.
  \ea
  \end{align}
\end{proposition}

\subsection{Ultra-discrete Toda lattice}

Suppose
\begin{align}
  V_i^t>0, \ I_i^t>0, \nn \\
  \prod_{i=1}^{g+1}V_i^t < \prod_{i=1}^{g+1}I_i^t. \label{IVcondition}
\end{align}
In the UD-limit $\lim_{\ve\to 0} \Log_\ve$ with
the scale transformation 
$I_i = \e^{-\frac{Q_i}{\ve}}, V_i = \e^{-\frac{W_i}{\ve}}$,
the system \eqref{d-toda2} becomes the UD-pToda lattice 
\eqref{UD-pToda}.
Simultaneously, the limit of the conserved quantities 
$c_i = \e^{-\frac{C_i}{\ve}}$
become 
\begin{align}\label{C_i}
  \begin{split}
  &C_{g} = \min[\min_{1\leq i \leq g+1} Q_i,\min_{1\leq i \leq g+1} W_i],
  \\ 
  &C_{g-1} = \min[\min_{1\leq i<j \leq g+1} (Q_i+Q_j),
                  \min_{1\leq i<j \leq g+1} (W_i+W_j),
                  \min_{1\leq i,j\leq g+1, j \neq i,i-1} (Q_i+W_j)],
  \\ 
  & \vdots 
  \\
  &C_0 = \min[\sum_{i=1}^{g+1} Q_i, \sum_{i=1}^{g+1} W_i], \\
  &C_{-1}= \sum_{i=1}^{g+1} (Q_i+ W_i), 
  \end{split}
\end{align}
which are preserved under (\ref{UD-pToda}) by construction.
From the assumption (\ref{IVcondition}), we have
\begin{align*}
\sum_{i=1}^{g+1}W_i^t > \sum_{i=1}^{g+1}Q_i^t.
\end{align*}
We can set $C_g = 0$ without loss of generality 
corresponding to $c_g = 1$.

Next, we reconstruct the tropical curve 
$\tilde{\Gamma}_C$ by the UD-limit
of the real part of the spectral curve $\gamma_c$.
We write $\gamma_{\mr}$ for the real part of $\gamma = \gamma_c$.
Then the image of the map 
$\Log^2 : ~\C^2 \to \R^2; ~ (x,y) \mapsto (\log |x|, \log |y|)$
of $\gamma_{\mr}$ is the rim of the amoeba of $\gamma$.

In taking the UD-limit of the equation 
\eqref{spectralcurve}  with the scale transformation 
$c_i = e^{-\frac{C_i}{\ve}}, |x| = e^{-\frac{X}{\ve}}$ and 
$|y| = e^{-\frac{Y}{\ve}}$,
we have the following:
\\
\noi (i) $x>0,y>0$ leads to a contradiction.\\
\noi (ii) $x<0,y>0$. We have
$$\Gamma_2:\left\{
\ba{ll}
\ba{l}
\min[2Y,C_{-1}, (g+1)X+Y, (g-1)X+Y+C_{g-1},\dots, Y+C_0]\\
=\min[gX+Y+C_g, (g-2)X+Y+C_{g-2},\dots, X+Y+C_1] 
\ea & \ (g:\mbox{odd})\\
\ba{l}
\min[2Y,C_{-1}, gX+Y+C_g, (g-2)X+Y+C_{g-2},\dots, Y+C_0]\\
=\min[(g+1)X+Y, (g-1)X+Y+C_{g-1},\dots, X+Y+C_1] 
\ea & \ (g:\mbox{even})
\ea\right.$$
\noi (iii) $x<0,y<0$. We have
$$\Gamma_3:\left\{
\ba{ll}
\ba{l}\min[2Y,C_{-1}, gX+Y+C_g, (g-2)X+Y+C_{g-2},\dots, X+Y+C_1]\\
=\min[(g+1)X+Y, (g-1)X+Y+C_{g-1},\dots, Y+C_0]
\ea & \ (g:\mbox{odd})\\
\ba{l}
\min[2Y,C_{-1}, (g+1)X+Y, (g-1)X+Y+C_{g-1},\dots, X+Y+C_1]\\
=\min[gX+Y+C_g, (g-2)X+Y+C_{g-2},\dots, Y+C_0]
\ea & \ (g:\mbox{even})
\ea\right.$$
\noi (iv) $x>0,y<0$. We have
$$\Gamma_4:
\min[2Y,C_{-1}]=\min[(g+1)X+Y, gX+Y+C_g,\dots, Y+C_0]
.$$

\begin{figure}[htbp]

\begin{center}
\unitlength=1.2mm
\begin{picture}(50,60)(0,-20)
\includegraphics*[width=160pt,height=100pt]{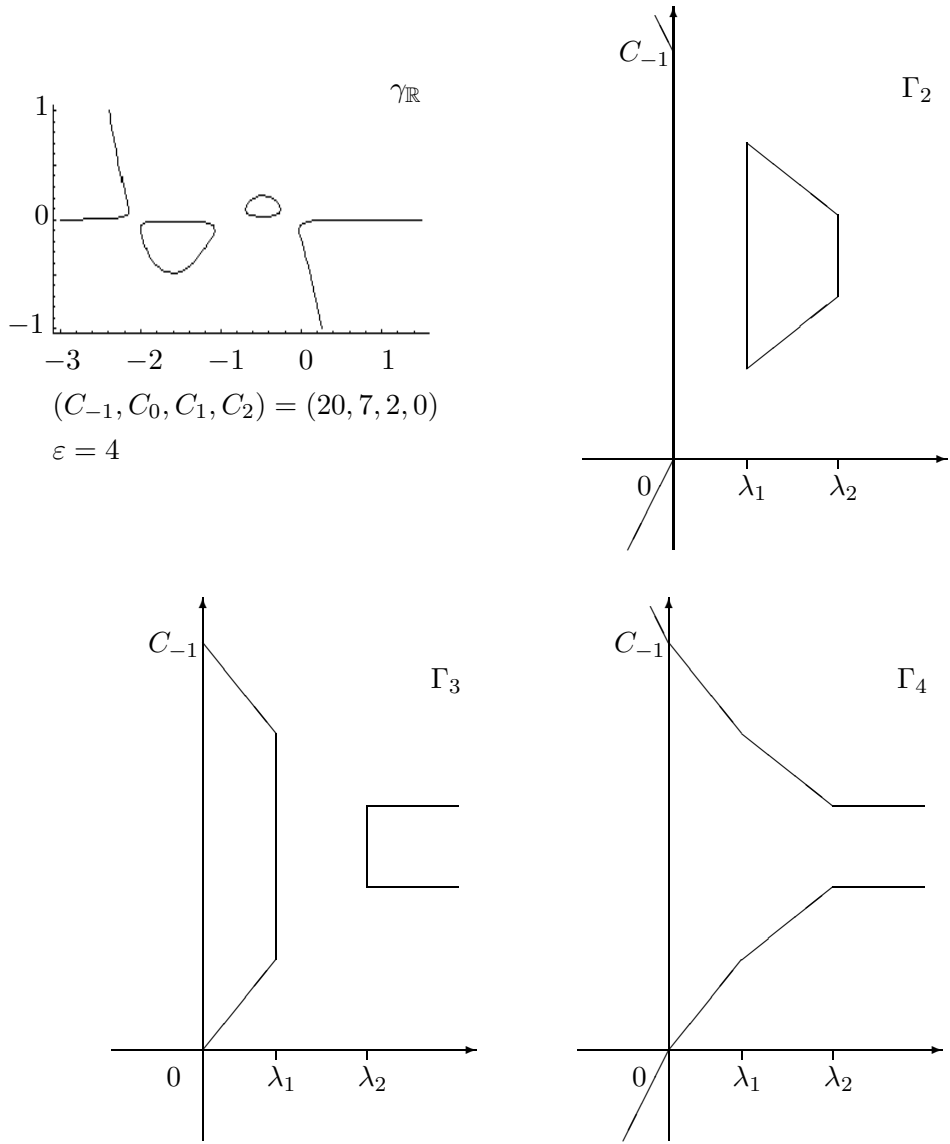}
\end{picture}
\begin{picture}(0,60)(50,-20)
\put(40,30){$\gamma_{\R}$}
%\put(3,-5){$Q_1=0,Q_2=2,Q_3=5$}
%\put(3,-10){$W_1=2,W_2=3,W_3=8,\ve=4$}
\put(3,-5){$(C_{-1},C_0,C_1,C_2) = (20,7,2,0)$}
\put(3,-10){$\ve=4$}
\put(2,0){$-3$}
\put(11,0){$-2$}
\put(20,0){$-1$}
\put(30,0){$0$}
\put(39,0){$1$}

\put(-2,4){$-1$}
\put(1,16){$0$}
\put(1,28){$1$}
\end{picture}
\begin{picture}(50,60)(0,0)
\put(45,50){$\Gamma_2$}
\put(20,10){\line(-1,0){10}}
\put(20,10){\vector(1,0){30}}% (20,10) the origin
\put(20,0){\line(0,1){10}}
\put(20,10){\line(0,1){45}}
\put(20,55){\vector(0,1){5}}

\put(14,54){$C_{-1}$}
\put(16,6){$0$}
\put(28,10){\line(0,-1){1}} % scale = 8picc
\put(27,6){$\lambda_1$}
\put(38,10){\line(0,-1){1}}
\put(37,6){$\lambda_2$}
\put(20,10){\line(-1,-2){5}}
%\put(20,10){\line(4,5){8}}
\put(28,20){\line(5,4){10}}

\put(20,55){\line(-1,2){2}}
%\put(20,75){\line(4,-5){8}}
\put(28,45){\line(5,-4){10}}

\put(28,20){\line(0,1){25}}
\put(38,28){\line(0,1){9}}

\end{picture}\\
\begin{picture}(50,65)(0,0)
\put(45,50){$\Gamma_3$}
\put(20,10){\line(-1,0){10}}
\put(20,10){\vector(1,0){30}}% (20,10) the origin
\put(20,0){\line(0,1){10}}
\put(20,10){\line(0,1){45}}
\put(20,55){\vector(0,1){5}}

\put(14,54){$C_{-1}$}
\put(16,6){$0$}
\put(28,10){\line(0,-1){1}} % scale = 8picc
\put(27,6){$\lambda_1$}
\put(38,10){\line(0,-1){1}}
\put(37,6){$\lambda_2$}

%\put(20,10){\line(-1,-2){5}}
\put(20,10){\line(4,5){8}}
%\put(28,20){\line(5,4){10}}
\put(38,28){\line(1,0){10}}

%\put(20,55){\line(-1,2){2}}
\put(20,55){\line(4,-5){8}}
%\put(28,45){\line(5,-4){10}}
\put(38,37){\line(1,0){10}}

\put(28,20){\line(0,1){25}}
\put(38,28){\line(0,1){9}}

\end{picture}
\begin{picture}(50,60)(0,0)
\put(45,50){$\Gamma_4$}
\put(20,10){\line(-1,0){10}}
\put(20,10){\vector(1,0){30}}% (20,10) the origin
\put(20,0){\line(0,1){10}}
\put(20,10){\line(0,1){45}}
\put(20,55){\vector(0,1){5}}

\put(14,54){$C_{-1}$}
\put(16,6){$0$}
\put(28,10){\line(0,-1){1}} % scale = 8picc
\put(27,6){$\lambda_1$}
\put(38,10){\line(0,-1){1}}
\put(37,6){$\lambda_2$}

\put(20,10){\line(-1,-2){5}}
\put(20,10){\line(4,5){8}}
\put(28,20){\line(5,4){10}}
\put(38,28){\line(1,0){10}}

\put(20,55){\line(-1,2){2}}
\put(20,55){\line(4,-5){8}}
\put(28,45){\line(5,-4){10}}
\put(38,37){\line(1,0){10}}

\end{picture}
\caption{Real and tropical curves}\label{RealTrop}
\end{center}
\end{figure}

Then we obtain the following.  
\begin{proposition}
For generic $C_i$'s which satisfy (\ref{CD-condition}), 
  $$\tilde{\Gamma}_C=\Gamma_2 \cup \Gamma_3
  =\Gamma_2 \cup \Gamma_3 \cup \Gamma_4$$ hold.
\end{proposition}
Fig.~\ref{RealTrop} shows $\gamma_{\R}$, $\Gamma_2$, $\Gamma_3$ and
$\Gamma_4$ in the case of $g=2$.

%%%%%%%%%%%%%%%%%%%%%%%%%%%%
\subsection{Eigenvector map}
%%%%%%%%%%%%%%%%%%%%%%%%%%%%

Let $\mathcal{T}$ be the phase space of the ultra-discrete
$(g+1)$-periodic Toda lattice, and $\mathcal{C}$ be the moduli 
space of the compact tropical curves $\Gamma_C$:
\begin{align*}
  &\mathcal{T} = \bigl\{(Q_1,\cdots,Q_{g+1},W_1,\cdots,W_{g+1}) 
                   \in \R^{2g+2} 
                   ~\big|~ \sum_{i=1}^{g+1} Q_i<\sum_{i=1}^{g+1} W_i 
                   \bigr\},
  \\
  &\mathcal{C} = \{C=(C_{-1},\cdots,C_{g-1},C_g) \} \simeq \R^{g+2}.
\end{align*}
We define a map $\Phi : ~ \mathcal{T} \to \mathcal{C}$ by (\ref{C_i}),
and set $\mathcal{T}_{C} = \Phi^{-1} (C)$ for $C \in \mathcal{C}$.

\begin{conjecture} \label{conj:Toda-iso}
For a generic $C=(C_{-1},\cdots,C_{g-1},C_g=0) \in \mathcal{C}$
which satisfies \eqref{CD-condition},
following are satisfied:
\\
{\rm (i)}  
$\mathcal{T}_{C} \simeq J(\Gamma_C)$.
\\
{\rm (ii)} 
Suppose $C \in \Z^{g+2}$, and let 
$(\mathcal{T}_{C})_{\Z}$ and $J_{\Z}(\Gamma_C)$ be 
the sets of lattice points in $\mathcal{T}_{C}$ and in $J(\Gamma_C)$ 
respectively.
Then the isomorphism of {\rm (i)} induces the bijection between
$(\mathcal{T}_{C})_{\Z}$ and $J_{\Z}(\Gamma_C)$. 
In particular, we have $|(\mathcal{T}_{C})_{\Z}| = \det \Lambda$.
\end{conjecture}

\begin{remark}
  This conjecture claims that 
  we need only a compact part $\Gamma_C$ of $\tilde{\Gamma}_C$
  to construct the isolevel set $\mathcal{T}_{C}$.
\end{remark}

In the rest of this section,
we construct the isomorphism $\pi: ~ \mathcal{T}_{C} 
\stackrel{\sim}{\longrightarrow} J(\Gamma_C)$
in the case of $g=1,2$ and $3$, 
by applying the technique of eigenvector map,
which is essentially the same with Sklyanin's separation 
of variable in our case 
(for example see \cite{Audin96, MM79}). 
The isomorphism $\pi$ is a composition of isomorphisms:
\begin{align*}
  \begin{matrix}
  \mathcal{T}_{C} & \stackrel{\psi}{\to} 
  & \mathcal{D}^g(\Gamma_C)
  & \stackrel{\eta}{\to} & J(\Gamma_C)  \end{matrix},
\end{align*}
where $\psi$ is called the eigenvector map (or separation of variables)
and $\eta$ is the Abel-Jacobi map \eqref{UD-AJ}. 

\begin{remark}
By concrete computation we also conjecture the following.
Define a translation operator $v$ as
$$
  v : J(\Gamma_C) \to J(\Gamma_C);~ z\mapsto z+
(\lambda_1,\lambda_2-\lambda_1,\dots,\lambda_g-\lambda_{g-1}).
$$
Then the following diagram is commutative:
$$
  \begin{matrix}
    \mathcal{T}_C  & \stackrel{\pi}{\longrightarrow} & J(\Gamma_C) \\
    \downarrow_{~T} & & \downarrow_{~v} \\
    \mathcal{T}_C  & \stackrel{\pi}{\longrightarrow} & J(\Gamma_C) \\ 
  \end{matrix}
$$
i. e. the flow of the UD-pToda is linearized on the tropical Jacobian.
It is easy to check it in the case of $g=1$.
\end{remark}

First we discuss the discrete case.
Let us consider the eigenvector $\phi$ of the Lax matrix $L^t(y)$.
Then $\phi$ is given by 
$$\phi= {}^t (f_1,f_2,\dots,f_{g},-f_{g+1}),$$
where $f_i (i=1,2,\dots,g)$ is
\begin{align*}
  f_i = \det
    \bordermatrix{
       &1       &2&  \cdots &  i          &  \cdots    &g \cr
      &l_{11}+x & l_{12} &\cdots & l_{1,g+1}     & \cdots&l_{1,g}\cr
       &l_{21}   & l_{22}+x & & l_{2,g+1}         & & l_{2,g}  \cr
    & \vdots      &  \vdots   & & \vdots    & & \vdots    \cr
    &l_{g,1}& l_{g,2}&\cdots & l_{g,g+1}   &\cdots  & l_{g,g}+x  
    }
\end{align*}
and 
$$
f_{g+1}=\det\begin{pmatrix}
      l_{11}+x & l_{12} &\cdots &l_{1,g}\cr
      l_{21}   & l_{22}+x &\cdots  & l_{2,g}  \cr
    \vdots      &  \vdots   &   & \vdots  \cr
   l_{g,1}& l_{g,2}&\cdots  & l_{g,g}+x  
    \end{pmatrix},
$$
where $l_{ij}=(L^t(y))_{ij}$.
The equation $f_{g+1}(x)=0 $ has the solution $x_1,x_2,\dots,x_g$,
each of which defines two points on $\gamma_c$: $(x_i,y_i), (x_i,y_i')$, 
where one of them (we assume that is $(x_i,y_i)$) 
leads $f_j=0$ for all $j$.
We choose $\{(x_i,y_i)\ |\  i=1,2,\dots,g \}$ or 
$\{(x_i,y_i')\ |\  i=1,2,\dots,g \}$ as a representative of $\pic^g(\gamma_c)$.
In the discrete case, this map induces an injection
$\mathcal{U}_c \hookrightarrow \pic^g(\gamma_c)$,
and the evolution equation \eqref{d-Toda} is linearized on 
the Jacobi variety of $\gamma_c$, 
$\mathrm{Jac}(\gamma_c) \simeq \pic^g(\gamma_c)$
(Cf. \cite{AdlerMoer80,KimijimaTokihiro02,MM79}).

%%%%%%%%%%%%%%%%%%%%%%%%%%%%%%
\subsection{The case of $g=1$}
%%%%%%%%%%%%%%%%%%%%%%%%%%%%%%

The Lax matrix is 
\begin{align*}
  L^t(y) = 
    \begin{pmatrix} 
      a_1 & 1-\frac{b_1}{y} \\
      b_2-y & a_2 
    \end{pmatrix} 
\end{align*}
and the conserved quantities are
\begin{align*} 
  c_{-1}=b_1 b_2,\ c_0=a_1a_2-b_1-b_2,\ c_1=a_1+a_2.
\end{align*}
When $f_2 = a_1+x = 0$, \eqref{spectralcurve} becomes 
\begin{align*}
  f(x,y) = (y-b_1)(y-b_2) = 0.
\end{align*}
Thus we define the map $\mathcal{U}_c \to \gamma_c$ by
$u^t \mapsto (x_1 = -a_1, y_1 = b_1)$.

In the ultra-discrete limit, the map $\psi: \mathcal{T}_C \to \Gamma_C$ 
is given by
$$
  (Q_1,Q_2,W_1,W_2) \mapsto (X_1 = \min[Q_2,W_1], Y_1 = Q_1+W_1)\in 
  \Gamma_2
$$
where $C_{-1} = Q_1+Q_2+W_1+W_2$, $C_0 = Q_1+Q_2$ and 
$C_1 = \min[Q_1,Q_2,W_1,W_2] = 0$. 
We see the following:
\begin{proposition}\label{eigenvm-g=1}
  The map $\psi$ is bijective.
  In particular, $\mathcal{T}_C \simeq J(\Gamma_C)$.
\end{proposition}

\begin{proof}
By construction it is obvious that the image of $\psi$ is included in 
$\Gamma_C$. Inversely, solving
\begin{align*}
&a_1=-x,  \  
b_1=y\\
& a_2=\frac{c_0+b_1+b_2}{a_1},  \  
b_2=\frac{c_{-1}}{y}
\end{align*}
for $I_i,V_j$, we have the solutions $(I_i,V_j)$ and $(I_i',V_j')$ 
($i,j=1,2$):
\begin{align*}
&I_1+I_1'=-\frac{c_0+2y}{x}, \quad
I_2+I_2'=-\frac{x(2c_{-1}+c_0y)}{c_{-1}+c_0y+y^2}\\&
V_1=\frac{y}{I_1},\ V_1'=\frac{y}{I_1'},\ 
V_2=\frac{c_{-1}}{yI_2},\ V_2'=\frac{c_{-1}}{yI_2'},
\end{align*}
where we assume $I_i\geq I_i'$.
Only $(I_i,V_j)$ satisfies the assumption (\ref{IVcondition}).
By the UD-limit, 
we have the inverse of $\psi$ as
\begin{align*}
Q_1 =& \min[C_0,Y]-X\\
Q_2 =& X+\min[C_{-1},C_0+Y]-\min[C_{-1},C_0+Y,2Y]\\
W_1 =& Y-Q_1\\
W_2 =& C_{-1}-Y-Q_2.
\end{align*}
\end{proof}

%%%%%%%%%%%%%%%%%%%%%%%%%%%%%%
\subsection{The case of $g=2$}
%%%%%%%%%%%%%%%%%%%%%%%%%%%%%%

In this and the next subsection we denote $\min[\quad ]$ simply by $[\quad ]$.
The Lax matrix is
\begin{align*}
  L^t(y) = 
    \begin{pmatrix} 
      a_1 & 1&\frac{b_1}{y} \\
      b_2 & a_2&1\\ 
      y & b_3&a_3 
    \end{pmatrix}, 
\end{align*}
and the conserved quantities are
\begin{align*} 
  c_{-1}&=b_1 b_2 b_3,\ c_0=a_1a_2a_3-a_2b_1-a_3b_2-a_1b_3,\\
  c_1&=a_1a_2+a_2a_3+a_3a_1-b_1-b_2-b_3,\ c_2=a_1+a_2+a_3 .
\end{align*}
The UD-pToda \eqref{UD-pToda} is
\begin{align}
\ba{rcl}
Q_i^{t+1}&=& [W_i^t,Q_i^t-X_i^t]\\
W_i^{t+1}&=&\disp Q_{i+1}^t+W_i^t-Q_i^{t+1}
\ea
\end{align}
with
$$X_i^t=[0,W_{i-1}^t-Q_{i-1}^t,W_{i-1}^t+W_{i-2}^t-Q_{i-1}^t-Q_{i-2}^t],$$
and the conserved quantities \eqref{C_i} become 
\begin{align*}
  C_2 =& [Q_1,Q_2,Q_3,W_1,W_2,W_3]=0\\
  C_1 =& [Q_1+Q_2,Q_2+Q_3,Q_3+Q_1,W_1+W_2,W_2+W_3,W_3+W_1,\\
       & \ Q_1+W_2, Q_2+W_3, Q_3+W_1]\\
  C_0 =& [Q_1+Q_2+Q_3, W_1+W_2+W_3]\\
  C_{-1}=& Q_1+Q_2+Q_3+ W_1+W_2+W_3 .
\end{align*}
The tropical spectral curve is the set sum of
$$\Gamma_2:
[2Y,C_{-1}, 2X+Y+C_2,Y+C_0]=[3X+Y,X+Y+C_1] $$
and
$$\Gamma_3:
[2Y,C_{-1}, 3X+Y,X+Y+C_1]=[2X+Y+C_2, Y+C_0].$$
The eigenvector of the Lax matrix is given by
$$f_1=\begin{vmatrix} 
      \frac{b_1}{y}&1 \\
      1& a_2+x 
    \end{vmatrix},\
  f_2=\begin{vmatrix} 
      a_1+x&\frac{b_1}{y} \\
      b_2& 1 
    \end{vmatrix},\   
  f_3=\begin{vmatrix} 
      a_1+x & 1 \\
      b_2 & a_2+x 
    \end{vmatrix}.$$
When $f_3=0$, \eqref{spectralcurve} reduces to
$$f(x,y)=(y-b_1(x+a_2))(y-b_3(x+a_1))=0.$$
The solutions are
\begin{align}\label{SoV-g=2}
  \begin{split}
  &x_1+x_2=-a_1-a_2,\ x_1x_2=a_1a_2-b_2\\
  &y_i=b_1(x_i+a_2),\ y_i'=b_3(x_i+a_1) ~\text{ for $i=1,2$}.
  \end{split}
\end{align}
For the UD-limit we use another expression of $y_i$:
$$y_i=\frac{c_{-1}}{b_3(x_i+a_1)}.$$

Under the assumption $x_1,x_2<0$, $y_1<0$, $y_2>0$ (for small $\ve>0$), 
the UD-limit of \eqref{SoV-g=2} leads:
\begin{align}\label{X_1X_2}
X_1 =& [Q_2, Q_3, W_1, W_2]\\
X_2 =& [Q_2+Q_3, W_1+W_2, Q_3+W_1] - X_1 \nn
\end{align}
and
\begin{align*}
Y_1 =& \left\{ \ba{ll}
Y_1^a:= Q_1+W_1+X_1 &\mbox{ if } X_1 < [Q_3, W_2]\\
Y_1^b:= C_{-1} - (Q_3+W_3+X_1)  &\mbox{ if } X_1 < [Q_2, W_1] 
\ea\right.\\
Y_2 =& \left\{ \ba{ll}
Y_2^a:= Q_1+W_1+[Q_3,W_2] &\mbox{ if } X_2 > [Q_3, W_2]\\
Y_2^b:= C_{-1} - (Q_3+W_3+[Q_2,W_1]) &\mbox{ if } X_2 > [Q_2, W_1] 
\ea\right..
\end{align*}

The following lemma can be proved elementarily.
\begin{lemma}\label{q2w1}
{\rm (i)} $C_2(=0) \leq X_1\leq C_2+\lambda_1 \leq X_2 \leq C_2+\lambda_2$.\\
{\rm (ii)} $X_1 = [[Q_2,W_1],[Q_3,W_2]],\ 
X_2 \geq \max[[Q_2,W_1],[Q_3,W_2]]$.\\
{\rm (iii)} If $[Q_2,W_1]=[Q_3,W_2]$, then \\
\hspace{5mm} {\rm (iii-1)}$X_1=X_2$ and thus $Y_1^a=Y_2^a$ and $Y_1^b=Y_2^b$\\
\hspace{5mm} or 
{\rm (iii-2)}
$Y_1^a = Y_1^b$ and $Y_2^a = Y_2^b$\\
hold.
\end{lemma}

By Lemma \ref{q2w1},
the correspondence between $(Q_1,Q_2,Q_3,W_1,W_2,W_3) \in \mathcal{T}_C$ 
and $(X_1,Y_1)+(X_2,Y_2) \in \Div_{\mathrm{eff}}^2(\Gamma_C)$  
is uniquely expanded as a continuous map
$\psi: \mathcal{T}_C \to \Div_{\mathrm{eff}}^2(\Gamma_C)$
by (\ref{X_1X_2}) and
\begin{align*}
\left.\ba{l}
Y_1 = Y_1^a= Q_1+W_1+[Q_2,W_1] \\
Y_2 = Y_2^b= C_{-1} - (Q_3+W_3+[Q_2,W_1])
\ea\right\}&
\mbox{ if } [Q_2,W_1] \leq [Q_3, W_2],\\
\left.\ba{l}
Y_1 = Y_1^b= C_{-1} - (Q_3+W_3+[Q_3,W_2]) \\
Y_2 = Y_2^a= Q_1+W_1+[Q_3,W_2] 
\ea\right\}& 
\mbox{ if } [Q_3,W_2] \leq [Q_2, W_1].
\end{align*}
(When $X_1=X_2$, we can exchange $Y_1$ and $Y_2$.)  

\begin{lemma}\label{X_1=X_2}
The image of $\psi$ is included in $\mathcal{D}^2(\Gamma_C)$, i.e.
if $X_1=X_2$, then $(X_1,Y_1)$ or $(X_2,Y_2)$ is at the end
point of $\alpha_{12}$.
\end{lemma}

\begin{proof}
By Lemma~\ref{q2w1}(ii), we have $[Q_2,W_1]=[Q_3,W_2]$.
Without loss of generality we can assume $Q_1=0$.
(i) $Q_2=Q_3\leq W_1,W_2$ leads $C_1=Q_2$ and $C_2=2Q_2$, which 
contradict to the smoothness (\ref{CD-condition}).
(ii) $Q_2=W_2< Q_3,W_1$ leads $X_1=Q_2$ and $X_2>Q_2$; 
which is a contradiction.
(iii) $W_1=Q_3\leq Q_2,W_2$ leads $C_1=W_1$ and $Y_1^a=2C_1$.
(iv) $W_1=W_2\leq Q_2,Q_3$ leads $C_1=W_1$ and $Y_1^a=2C_1$.
\end{proof}

Inversely, solving
\begin{align*}
a_1=-\frac{x_1y_1-x_2y_2}{y_1-y_2}, & \  
b_1=\frac{y_1-y_2}{x_1-x_2}\\
a_2=\frac{x_1y_2-x_2y_1}{y_1-y_2}, & \  
b_2=-\frac{y_1y_2(x_1-x_2)^2}{(y_1-y_2)^2}\\
a_3=\frac{c_0a_1b_3+a_2b_1}{a_1a_2-b_2}, & \ 
b_3=-\frac{c_{-1}(y_1-y_2)}{y_1y_2(x_1-x_2)}
\end{align*}
for $I_i,V_j$, 
we have (e.g.)
\begin{align*}
I_1+I_1'=\frac{c_0(x_1-x_2)+2(x_1y_2-x_2y_1)}{
x_1x_2(x_1-x_2)}.
\end{align*}
By the UD-limit, 
we have the inverse of $\psi$ if $X_1<X_2$
\begin{align*}
Q_1 =& [C_0+X_1, U_2] - (2X_1+X_2)\\
Q_2 =& 2X_1+[C_{-1}+U_1,Y_1+Y_2+U_2,
C_0+[X_1+Y_1+Y_2, X_2+2[Y_1, Y_2]] ] \\
&- [Y_1,Y_2] - [C_{-1}+2X_1, C_0+X_1+U_2, 2U_2]\\
Q_3 =& X_1+X_2+[Y_1, Y_2]+[C_{-1}+U_1, C_0 +X_1+Y_1+Y_2] \\&
 -  [C_{-1}+2U_1, C_0+X_1+Y_1+Y_2+U_1, 2X_1+2Y_1+2Y_2]\\
W_1&=[Y_1,Y_2]-X_1-Q_1\\
W_2&= Y_1+Y_2+2X_1-2[Y_1,Y_2]-Q_2\\
W_3&=[C_{-1}+[Y_1,Y_2]-Y_1-Y_2-X_1-Q_3
\end{align*}
with
\begin{align*}
  U_1=[X_1+Y_1, X_2+Y_2], \qquad
  U_2=[X_1+Y_2, X_2+Y_1].
\end{align*}
By Lemma~\ref{X_1=X_2}, the inverse is uniquely expanded
as a continuous map to the case of $X_1=X_2$.

Now we have the following.

\begin{proposition}\label{eigenvm-g=2}
  The UD-eigenvector map $\psi:\mathcal{T}_C \to \mathcal{D}^2(\Gamma_C)$
  is bijective.
\end{proposition}

%%%%%%%%%%%%%%%%%%%%%%%%%%%%%%
\subsection{The case of $g=3$}
%%%%%%%%%%%%%%%%%%%%%%%%%%%%%%

In the case of $g=3$ we present the ultra-discrete 
eigenvector map 
$\psi: \mathcal{T}_C \to \mathcal{D}^3(\Gamma_C)$.
However, for the reason of complexity, 
we will omit to present the inverse mapping and to prove the bijectivity.

The solutions of $f_4=0$ and $f(x,y)=0$ are
\begin{align}\label{SoV-g=3}
\begin{split}
  &x_1+x_2+x_3=-a_1-a_2-a_2\\
  &x_1x_2+x_2x_3+x_3x_1=a_1a_2+a_2a_3+a_3a_1-b_2-b_3 \\
  &x_1x_2x_3=-a_1a_2a_3+a_1b_3+a_3b_2 \\
  &y_i=b_1((a_2+x_i)(a_3+x_i)-b_3)), ~~y_i'=b_3 ((a_1+x_i)(a_2+x_i)-b_2))
    ~\text{ for $i=1,2,3$}.
\end{split}
\end{align}
For the UD-limit we use other expressions of $y_i$:
%$$y_i=b_1b_2\frac{a_3+x_i}{a_1+x_i}
% =\frac{c_{-1}}{b_4 ((a_1+x_i)(a_2+x_i)-b_2))}.$$
$$y_i=b_1b_2\frac{a_3+x_i}{a_1+x_i}
 =\frac{c_{-1}}{b_4 ((a_1+x_i)(a_2+x_i)-b_2))}.$$

The UD-limit of \eqref{SoV-g=3} leads the UD-eigenvector map 
$\psi:\mathcal{T}_C \to \mathcal{D}^3(\Gamma_C)$:
\begin{align*}
  X_1 =& [Q_2, Q_3, Q_4, W_1, W_2, W_3]\\
  X_2 =& [Q_2 + Q_3, Q_3+Q_4, Q_2+Q_4,
   W_1+W_2, W_2+W_3, W_1+W_3, Q_4+W_1, \\& Q_4+W_2, Q_2+W_3, Q_3+W_1] - X_1\\
  X_3 =& [Q_2+Q_3+Q_4, W_1+Q_3+Q_4, W_1+W_2+Q_4, W_1+W_2+W_3] - (
          X_1+X_2)\\
  Y_i=&Y_i^{s_i}  ~\text{ for $i=1,2,3$},
\end{align*}
where
\begin{align*}    
    Y_i^1 =& Q_1+W_1+[2X_i, X_i+[Q_3, Q_4, W_2, W_3], [Q_3+Q_4, 
          W_2+W_3,Q_4+W_2]]\\
    Y_i^2 =& Q_1+W_1+Q_2+W_2+[Q_4, W_3, X_i] - [Q_2, W_1, X_i]\\
    Y_i^3 =& C_{-1} - (Q_4+W_4+[2X_i, X_i+[Q_2, 
          Q_3, W_1, W_2], [Q_2+Q_3, W_1+W_2,Q_3+W_1]])
\end{align*}
and  $s_i$ is defined as follows.\\
(i) Set $A^1,A^2,A^3,B^1,B^3$ as 
$A^1=[Q_2,W_1], A^2=[Q_3,W_2], A^3=[Q_4,W_3]$,\\ 
$B^1=[Q_3+Q_4,W_2+W_3,Q_4+W_2],
B^3=[Q_2+Q_3,W_1+W_2,Q_3+W_1]$, and define $s_1$ by\\
$s_1=1$ if $A^1\leq [A^2,A^3]$,\\
$s_1=2$ if $A^2\leq [A^3,A^1]$,\\
$s_1=3$ if $A^3\leq [A^1,A^2]$.\\
If $s_1$ has two or more possibilities, choose one of them.\\
(ii) Define $s_2$ and $s_3$ so that $s_i\neq s_j (i,j=1,2,3)$ by\\ 
$s_2=1$ if  $X_2+[A^2,A^3]<B^1$,\\
$s_2=2$ if  $A^1<X_2<A^3$ or $A^3<X_2<A^1$,\\
$s_2=3$ if  $X_2+[A^1,A^2]<B^3$\\
and\\
$s_3=1$ if  $X_3+[A^2,A^3]>B^1$,\\
$s_3=2$ if  $X_3>\max[A^1,A^3]$,\\
$s_3=3$ if  $X_3+[A^1,A^2]>B^3$.\\
(iii) If both $s_2$ and $s_3$ are not determined by (ii), then
 choose $s_2$ and $s_3$ arbitrarily under keeping $s_i\neq s_j (i,j=1,2,3)$.

%%%%%%%%%%%%%%%%%%%%%%%%%%%%%%%%%%%%%%%%%%%%%%%%%%
\section{From the UD-pToda to the pBBS}
%%%%%%%%%%%%%%%%%%%%%%%%%%%%%%%%%%%%%%%%%%%%%%%%%%
%%%%%%%%%%%%%%%%%%%%%%%%%%%%%%%%%%%%%%%%%%%%%
\subsection{The structure of $\mathcal{T}_C$}
%%%%%%%%%%%%%%%%%%%%%%%%%%%%%%%%%%%%%%%%%%%%%

Fix a generic $C \in \mathcal{C}$ with $C_g = 0$.
Define a shift operator
$s :~ \mathcal{T}_C \to \mathcal{T}_C;$
\begin{align}\label{shift-s} 
  (Q_1,\cdots,Q_{g+1},W_1,\cdots,W_{g+1}) \mapsto 
  (Q_2,\cdots,Q_{g+1},Q_1,W_2,\cdots,W_{g+1},W_1).
\end{align}
Note $s^{g+1} = id$.
We define a subspace $T^0_{C}$ of $\mathcal{T}_{C}$:
\begin{align}
  \label{T^0}
  T^0_{C} = \bigl\{ (Q_1,\cdots,Q_{g+1},W_1,\cdots,W_{g+1})
                              \in \mathcal{T}_{C} ~|~ 
                   \text{(a) $W_1 > 0$, and 
                         (b) $Q_1 = 0$ or $W_{g+1}=0$}. \bigr\}.  
\end{align}
We write $T^i_{C}$ for the set given by
$$
  T^i_{C} = \{ s^i (\tau) ~|~ \tau \in T^0_{C} \},
  ~~ \text{ for $i \in \Z$.}
$$

\begin{proposition}
  \label{prop:sT}
  {\rm (i)} $T^i_{C} \cap T^j_{C} = \emptyset$
       for $i \neq j \mod g+1$,
  \qquad
  {\rm (ii)} $\displaystyle{\mathcal{T}_{C} = \bigcup_{i=0}^g T^i_{C}}$.
\end{proposition}
First we show 
\begin{lemma}
  \label{lemma:T^0}
  If $\tau = (Q_1,\cdots,Q_{g+1},W_1,\cdots,W_{g+1}) \in T^0_{C}$, 
  then $Q_i > 0$ for $2 \leq i \leq g$, and 
  $W_j > 0$ for $1 \leq j \leq g$.
\end{lemma}
\begin{proof}
  Recall that the conserved quantity $C_{g-1}$ \eqref{C_g-1} 
  for $\mathcal{T}_{C}$ satisfies $C_{g-1} > 0$.
  For $\tau = (Q_1,\cdots,Q_{g+1},W_1,\cdots,W_{g+1}) \in \mathcal{T}_{C}$,
  the following properties (b1) and (b2) hold:
  \\
  (b1) When $Q_1 = 0$, we have 
  \begin{align}
    \label{C-Q=0}
  C_{g-1} = \min[\min_{2 \leq i \leq g+1} Q_i, \min_{2 \leq i \leq g} W_i,
                 W_1+W_{g+1}] > 0.   
  \end{align}
  Thus we obtain $Q_i > 0$ for $2 \leq i \leq g+1$
  and $W_j > 0$ for $2 \leq i \leq g$. 
  \\
  (b2) When $W_{g+1} = 0$, we have
  \begin{align}
    \label{C-W=0}
  C_{g-1} = \min[\min_{2 \leq i \leq g} Q_i, \min_{1 \leq i \leq g} W_i,
                 Q_1+Q_{g+1}] > 0.   
  \end{align}
  Thus we obtain $Q_i > 0$ for $2 \leq i \leq g$
  and $W_j > 0$ for $1 \leq i \leq g$.

  If we further assume $\tau \in T^0_{C}$, 
  we have $W_1 > 0$, and (b1) or (b2) is satisfied.
%  (In particular, when (b2) holds, $\tau$ turns out to belong to $T^0_{C}$ 
%  without the assumption.)
  Thus one obtains the claim.
\end{proof}

\begin{proof}(Proposition \ref{prop:sT})
  \\
  (i) 
  Note that 
  $$
  T^i_{C} = 
  \{(Q_1,\cdots,Q_{g+1},W_1,\cdots,W_{g+1}) ~|~ 
  \text{(a) $W_{i+1} > 0$, and (b) $Q_{i+1} = 0$ or $W_i = 0$} \}.
  $$
  We check that 
  if $\tau = (Q_1,\cdots,Q_{g+1},W_1,\cdots,W_{g+1}) \in T^0_{C}$ 
  then it satisfies (a') $W_{i+1} = 0$, or (b') $Q_{i+1} > 0$ and $W_i > 0$,
  for $i=1,\cdots,g$.
  For $i=1,\cdots, g-1$, (b') is satisfied due to Lemma \ref{lemma:T^0}. 
  For $i=g$, (b') is satisfied when $Q_1=0$ and 
  (a') is satisfied when $W_{g+1}=0$.
  \\
  (ii)
  Is is trivial that 
  $\displaystyle{\mathcal{T}_{C} \supset \bigcup_{i=0}^g T^i_{C}}$.
  We show
  $\displaystyle{\mathcal{T}_{C} \subset \bigcup_{i=0}^g T^i_{C}}$.
  Since $C_g = 0$,
  for $\tau \in \mathcal{T}_{C}$ we assume $Q_1 = 0$ or $W_{g+1} = 0$
  without loss of the generality.
  When $Q_1 = 0$, \eqref{C-Q=0} denotes 
  $Q_2,\cdots,Q_{g+1},W_2,\cdots,W_g > 0$ and $W_1+W_{g+1} > 0$.
  Thus we see $\tau \in T^1_{C}$ when $W_1 = 0$, 
  and $\tau \in T^0_{C}$ when $W_1 > 0$.
  In the same way,
  when $W_{g+1} = 0$ it is easy to see that  
  \eqref{C-W=0} indicates $\tau \in T^0_{C}$. 
\end{proof}

\subsection{Periodic BBS}

Fix $L \in \Z_{>0}$.
The $L$-periodic box-ball system is a cellular automaton
that the finite number of balls move in a periodic array of $L$ boxes,
where each box has one ball at most \cite{YuraTokihiro02}.
We assume that the number of balls $|\lambda|$ satisfies
$2 |\lambda| < L$.
The time evolution of the pBBS from the time step $t$ to $t+1$
is given as follows:
\begin{enumerate}
  \item
  Choose one ball and move it to the leftmost empty box to its right.
  \item
  Choose one of unmoved balls and
  move it as (1),
  ignoring the boxes to which and from which the balls were moved 
  in this time step.
  \item
  Continue (2) until every ball moves once.
\end{enumerate}
This system has conserved quantities parametrized by
a non-decreasing array
$\lambda = (\lambda_1,\cdots,\lambda_g) \in (\Z_{> 0})^g$
such that $\sum_{i=1}^g \lambda_i = |\lambda|$ for some $g \in \Z_{>0}$.
In the following we write $0$ and $1$ for 
``an empty box" and ``an occupied box" respectively,
and let $B_L \simeq \{0,1\}^{\times L}$ 
be the phase space of $L$-periodic BBS.
We show examples of the evolution of $b(t) \in B_L$ 
as time $t$ goes:
\begin{example}\label{mapbeta}
  The case of (i) $(L,\lambda_1) = (8,3)$ and 
  (ii) $(L,\lambda_1,\lambda_2) = (7,1,2)$:
  \begin{align*}
    \begin{matrix}
    \text{(i)} \\[1mm]
    t & b(t) \\
    0 & 00111000 \\
    1 & 00000111 \\
    2 & 11100000 \\
    3 & 00011100 \\
    4 & 10000011 \\
    5 & 01110000
    \end{matrix}
    \hspace{2cm}
    \begin{matrix}
    \text{(ii)} \\[1mm]
    t & b(t)    & & \beta(b(t)) & &  T^t (\beta(b(0)))\\
    0 & 0100110 & & (0,1,2,1,2,1)& & (0,1,2,1,2,1) \\ 
    1 & 1010001 & &(1,1,1,1,3,0) & & (1,1,1,1,3,0) \\
    2 & 0101100 & & (0,1,2,1,1,2)& & (1,2,0,1,2,1) \\
    3 & 0010011 & & (0,1,2,2,2,0)& & (1,2,0,2,0,2) \\
    4 & 1101000 & & (2,1,0,1,3,0)& & (1,0,2,3,0,1) \\
    5 & 0010110 & & (0,1,2,2,1,1)& & (2,0,1,1,2,1)
    \end{matrix}
  \end{align*}
\end{example}
Roughly speaking, $g$ is the number of consecutive clusters of $1$'s,
and $(\lambda_1,\cdots,\lambda_g)$ corresponds to 
the number of $1$'s in each cluster. 

The injection from $B_L$ to $\mathcal{T}$ is introduced in 
\cite{KimijimaTokihiro02}.
Fix a generic $C \in \mathcal{C} \cap \Z^{g+2}$ which satisfies 
\eqref{CD-condition}
with $C_{-1} = L$ and $C_g = 0$,
and set $\lambda = (\lambda_1,\cdots,\lambda_g)$ \eqref{partition}.
Note that the generic condition for $C$ corresponds
to the condition: $0<\lambda_1 < \lambda_2<\cdots< \lambda_g$.
Let $B_{L,\lambda} \subset B_L$ be a set of the states 
whose conserved quantity is $\lambda$. 
Then the injection 
$\beta: ~B_{L,\lambda} \hookrightarrow (\mathcal{T}_C)_{\Z};
~ b \mapsto (Q_1,\cdots,Q_{g+1},W_1,\cdots,W_{g+1})$ is defined as follows:
\begin{enumerate} 
\item 
if the leftmost entry of $b$ is $1$, then set 
$Q_1 = \sharp(\text{the first consecutive $1$'s from the left})$,
otherwise set $Q_1 = 0$.

\item
Set $W_i = \sharp(\text{the $i$-th consecutive $0$'s from the left})$
for $i=1,\cdots, g+1$.
If $Q_1 \neq 0$,
set $Q_i = \sharp(\text{the $i$-th consecutive $1$'s from the left})$,
otherwise set 
$Q_i = \sharp(\text{the ($i-1$)-th consecutive $1$'s from the left})$
for $i=2,\cdots, g+1$.
\end{enumerate}

\begin{proposition}
  \label{conj:T/S}
  $\beta : ~ B_{L,\lambda} \to  
  (T^0_{C})_{\Z} := T^0_{C} \cap \Z^{2(g+1)}$ is a bijection.
  In particular, we have the bijection between
  $(\mathcal{T}_C)_{\Z} / \{ \tau \sim s(\tau) 
   ~|~ \tau \in (\mathcal{T}_C)_{\Z} \}$ and $B_{L,\lambda}$,
  which leads to $|(\mathcal{T}_C)_{\Z}| = (g+1) |B_{L,\lambda}|$.
\end{proposition}

\begin{proof}
  By the definition of the map $\beta$,
  it is obvious $\beta(B_{L,\lambda}) \subset (T^0_{C})_{\Z}$.
  From Lemma \ref{lemma:T^0}, each $\tau \in T^0_{C}$
  gives the array $(Q_1,W_1,Q_2,\cdots,W_g,Q_{g+1},W_{g+1})$
  where $W_1,Q_2,\cdots,W_g > 0$.  
  We define a map $\rho : ~(T^0_{C})_{\Z}
  \to B_{L,\lambda}$ as follows:
  when $Q_1 = 0$, set $\rho(\tau)$ as
  $$
    \underbrace{0 \cdots 0}_{W_1} \underbrace{1 \cdots 1}_{Q_2}
    ~~\cdots~~ 
    \underbrace{1 \cdots 1}_{Q_{g+1}} \underbrace{0 \cdots 0}_{W_{g+1}}
  $$
  where $W_{g+1}$ can be zero.
  When $W_{g+1} = 0$, set $\eta(\tau)$ as
  $$
    \underbrace{1 \cdots 1}_{Q_1} \underbrace{0 \cdots 0}_{W_1}
    ~~\cdots~~ 
    \underbrace{0 \cdots 0}_{W_{g}} \underbrace{1 \cdots 1}_{Q_{g+1}}
  $$
  where one of $Q_1$ and $Q_{g+1}$ can be zero.
  In both cases, it is clear that $\beta \cdot \rho(\tau) = \tau$.
  Thus $\rho = \beta^{-1}$. 
\end{proof}

From Prop.~\ref{prop:sT} and \ref{conj:T/S} we can put back
$b(t)$ from the solution of the UD-pToda lattice with the initial state 
$\beta(b(0))$ (see Example~\ref{mapbeta} (ii)).

Lemma~\ref{lemma:Lambda-A} clarifies the algebro-geometrical 
meaning of $J'(\Gamma_C)$ \eqref{Mat-T}
which was first introduced in the study of the pBBS by Kuniba et al:
\begin{theorem} \cite[Theorem 3.11]{KTT06}
  Let $J'_{\Z}(\Gamma_C)$ 
  be the set of lattice points in $J'(\Gamma_C)$. 
  Then the bijection between 
  $B_{L,\lambda}$ and $J'_{\Z}(\Gamma_C)$ is induced by
  Kerov-Kirillov-Reshetikhin bijection.
\end{theorem}
 
In the following diagram we summarize
the relation among the UD-pToda, the pBBS and 
the tropical Jacobian:
\begin{align}\label{BTJ-diagram}
  \begin{matrix} 
  B_{L,\lambda} & \stackrel{\beta}{\hookrl}_{_{/s}} & 
  (\mathcal{T}_C)_{\Z} & \subset & \mathcal{T}_C
  \\
  \downarrow \wr & & \downarrow \wr & & \downarrow \wr                      
  \\
  J'_{\Z}(\Gamma_C) & {\leftarrow}_{_{/\nu_{\Gamma}}} & J_{\Z}(\Gamma_C)
  & \subset & J(\Gamma_C)  
  \end{matrix}
\end{align}
Here {\small $/s$} and {\small $/\nu_{\Gamma}$} are the quotient maps
respectively induced by the shift operators $s$ \eqref{shift-s} 
and $\nu_{\Gamma}$ at Lemma \ref{lemma:Lambda-A} (ii).
The isomorphism of the right two downward maps are
the claim in Conjecture \ref{conj:Toda-iso}.
The diagram also indicates 
$J'(\Gamma_C) \simeq 
J(\Gamma_C)/\{ P \sim s^\ast(P) ~|~ P \in J(\Gamma_C)\}$.

%%%%%%%%%%%%%%%%%%%%%%%%%%%%%%%%%%%%%%%%

\end{document}